# A Direct Derivation of the Griffith-Irwin Relationship using a Crack tip Unloading Stress Wave Model.

C.E. Neal-Sturgess. Emeritus Professor of Mechanical Engineering, The University of Birmingham, UK.

## Abstract
It is the purpose of this short paper to show that the well known Griffith's theory of brittle fracture can be re-interpreted as a simple case of a modified Hamiltonian, and corresponds to a stationary action solution using the Euler-Lagrange equations. In this derivation it is still necessary to employ the equivalence with Irwin's stress intensity factor approach to obtain the Griffith-Irwin equation.
It is then shown that if physics of fracture is modelled using a simplified crack tip stress wave unloading model, then the Griffith-Irwin relationship can be derived in a single process, which is a major simplification of the procedure. Also the simplified crack tip stress wave unloading model employed gives a much clearer view of the basic physics of fracture.

## Introduction
The theory of elastic fracture originated by AA Griffith's in 1921 (Griffith 1921) is well known, and although Griffith's used the "*theorem of minimum potential energy*" (his italics) the manner in which he used this theorem seems to have obscured its more general usage. If it is set within its classical mechanics roots, it is clear that Griffith's used a restricted form of a modified Hamiltonian, as shown below. Also, as is widely known, in conventional treatments it is then necessary to equate Griffith's equation with that of Irwin (Irwin 1957) to get the usual Griffith–Irwin relationship.

In advanced texts on fracture Freund (Freund 1990), Kanninen and Popelar (Kanninen and Popelar 1985), and William's (Williams 1984) cracks are treated as dynamic entities, and the role of stress waves is clearly articulated, also see Kolsky (Kolsky 1963). However, in most non-advanced texts on fracture the role of stress waves are not included, leading to a possible misunderstanding of the fundamentals of fracture. The position adopted here is that all cracks are dynamic entities, driven by a crack tip stress wave unloading mechanism.

What is done here is to extend Griffith's approach by setting it within the concept of Hamiltonian mechanics (Stationary Action), and introducing a quasi-static stress wave unloading model for slow cracks, which connects the energy release mechanism with the 'K' dominated stress field at the tip of the crack.

**Monotonic fracture for a perfectly elastic material - The Griffith's Crack**

Consider the "Action", defined as:

$$S \equiv \int_{t_1}^{t_2} L(x, \dot{x}, t) \tag{1}$$

Where $L$ is the Lagrangian. Consider a function $x(t)$, for $t_1 < t < t_2$, which has fixed endpoints but is otherwise arbitrary. A function $x(t)$ which yields a stationary value of $S$ satisfies the Euler-Lagrange equation (Goldstein 1969):

$$\frac{d}{dt}\left(\frac{\partial L}{\partial \dot{x}}\right) - \frac{\partial L}{\partial x} = 0 \tag{2}$$

Hamiltonian mechanics provides an extremely powerful framework in modern physics (Penrose 2005), and for mechanical systems the Lagrangian is usually expressed as:

$$L = T - V \tag{3}$$

Where     $V$ = the gravitational potential energy
              $T$ = the kinetic potential energy

If crack is embedded, and the boundaries of the plate are at infinity, then the system is "closed", that is $\frac{dU_T}{dt} = 0$, and so non-extended Hamiltonian mechanics apply (Goldstein 1969). For the fracture problem if the crack is planar and propagating horizontally, at velocities which do not involve a consideration of crack inertia (Kanninen and Popelar 1985; Freund 1990) it can be assumed to have a constant value of the gravitational potential energy, and so the energy transfer is the strain energy stored in the body being converted into new crack surfaces. For this problem a modified Lagrangian can be defined as:

$$L = H - M - \Omega \tag{4}$$

Where:     $H$ = the energy required to create new crack surfaces
              $M$ = the strain energy released from the body
              $\Omega$ = the strain energy of the applied loads (taken to be zero for an infinite plate)

this technique is frequently used in solid mechanics Richards (Richards 1977) and Goldstein (Goldstein 1969). Assuming that a crack will propagate when the Euler-Lagrange equation is satisfied (i.e. that the Action is stationary), let $x = a$, where $a$ is the semi crack length, see Fig1. For a Griffith crack in an infinite body of negligible thickness in the z direction, and loaded by a remote stress $\sigma$ in the y direction. The energy required to create the crack surfaces per unit thickness can be defined as:

$$H = 2\gamma a \tag{4}$$

Where $\gamma$ is the energy to create new crack surfaces, misleadingly called the "surface energy" in many texts.

Griffith's assumed that when an idealised crack of length 2a is inserted into a stressed body a circular zone shown in Fig.1 is unloaded. Assuming that M is the strain energy released from the body by this unloading, and using the concept of energy = (energy per unit volume) x (the volume) gives:

$$M = \frac{\sigma^2}{2E}(vol) = \frac{\sigma^2}{2E}\left(\pi a^2\right) \tag{5}$$

Where $\sigma$ is the remote stress, and utilising the assumption for the area of the unloaded zone is a circular area of a radius equal to the crack semi length from (Benham, Armstrong et al. 1987), which a modern interpretation of the more complex assumptions Griffith's made, Fig. 1. This is clearly an idealised model, whereas in reality cracks propagate from the tip of the crack.

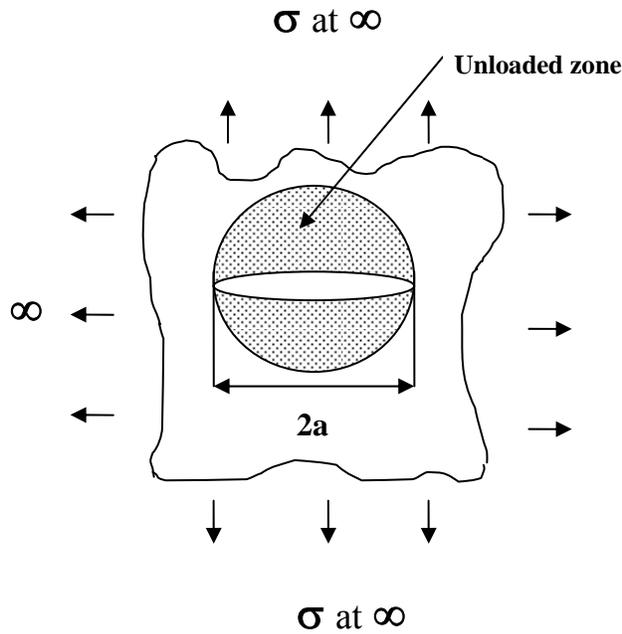

Fig. 1. The Griffith Crack

Hence $L = 2\gamma a - \dfrac{\sigma^2}{2E}\left(\pi a^2\right)$ (6)

Now as $\dfrac{d}{dt}\left(\dfrac{\partial L}{\partial a}\right) = 0 \therefore \dfrac{\partial L}{\partial a} = 0$ which is the form Griffith's used, and so

$$2\gamma E = \sigma^2 \pi a \tag{7}$$

Which is the Griffiths energy balance equation.

Irwin (Irwin 1957) derived a crack growth equation from considering the stress field ahead of the crack to be dominated by the stress intensity factor 'K', Fig 2 and considering crack closure.

Where $\quad K = \sigma\sqrt{\pi a}$

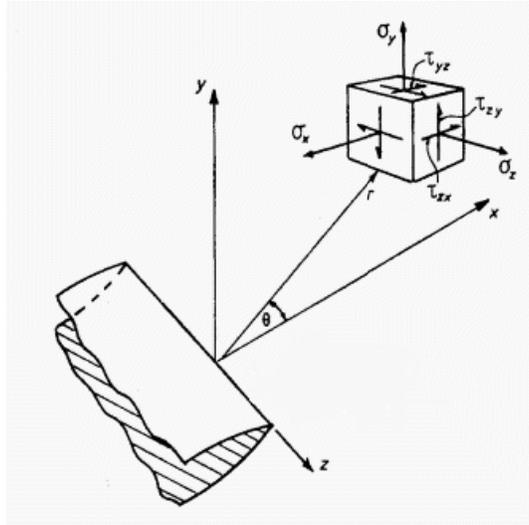

Fig. 2. Stress state at the crack tip

Irwin defined the term "strain energy release rate" (G) (Irwin 1957), and showed that, as all the stress components are functions of K, and as strain energy per unit volume is a function of $\sigma^2$, then:

$$G = 2\gamma E \propto K^2 \tag{8}$$

Therefore equating (7) and (8) gives

$$2\gamma E = \sigma^2 \pi a = K^2 \tag{9}$$

Which is the well known Griffith-Irwin relationship.

This relationship has been derived here using a Lagrangian to give the same result as Griffiths.

**A crack tip stress wave unloading model for a perfectly brittle elastic material:**

In the Griffith's model outlined above, current interpretations (Benham, Armstrong et al. 1987) of the area of the unloaded zone consider it to be a circular area equal in diameter to the length of the embedded crack i.e. 2a, which is a considerable idealisation, and does not bear any relevance to the actual physics of the situation. In the Griffith's idealisation this crack appears instantaneously in the loaded plate.

Considering the physics of the situation, in the loaded body the crack tip is under a tensile local stress $\sigma_{ij}$, where the distribution for the stress at the crack tip as shown in Fig 2. When the tip of an idealised atomically sharp crack separates the atomic bond at the crack tip separates and stress at the crack tip falls to zero. This "stress drop" as it is known in Tectonophysics can be up to around 70% of the stress at the crack tip (Loo, Yuan et al. 1990). As it is an unloading stress wave it may be modelled as a superimposed compressive wave which propagates outwards from the crack tip. This phenomena was filmed in 1959 by Schardin (Schardin 1959) using high speed photography, as shown in Fig.3, and by Theocaris and Georgiadis in 1984 (Theocaris and Georgiadis 1984). In the event of the crack tip stress falling to zero three types of waves are emitted, there are longitudinal (dilatational waves), shear (distortional) waves and Rayleigh surface waves (Kolsky 1963). The waves shown in Fig.3. are actually Raleigh surface waves, however Theocaris and Georgiadis also photographed longitudinal waves (Theocaris and Georgiadis 1984). In thin specimens both the longitudinal and surfaces waves propagate with circular wave fronts. In brittle materials under plane strain conditions Rayleigh waves do not propagate due to the lack of a plastic zone at the crack tip (Theocaris and Georgiadis 1984). It is these unloading waves which release the strain energy from the body, and makes it available for generating new crack surfaces. In a real solid, which is not infinite in extent, the longitudinal stress waves propagate across the body and when they reach a free surface they are virtually totally internally reflected as tensile waves. These tensile wave then propagates back across the body, reload the crack tip, and may cause another separation which generates another unloading wave (Kanninen and Popelar 1985; Freund 1990).

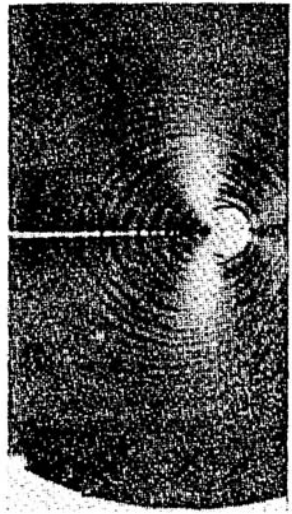

Fig.3 Dynamically Propagating Crack. From (Schardin 1959)

If the material is perfectly elastic, and the stress field at the tip of the crack is K dominated, then after Irwin the local stress state at the crack tip can be described as shown in equation 10. However for this plane stress idealisation of an infinite plate of negligible thickness only $\sigma_y$ is of significance i.e. $\sigma_{yy} = \sigma_y, \sigma_{xx} = \sigma_{zz} = 0$, hence:

$$\sigma_{ij} = \frac{K}{\sqrt{2\pi r}} f(\theta) = \sigma_y \qquad (10)$$

Letting $f(\theta) = 1$, then $r = \frac{K^2}{2\pi \sigma_y^2}$

Modelling the crack tip stress wave unloading as shown in Fig. 4:

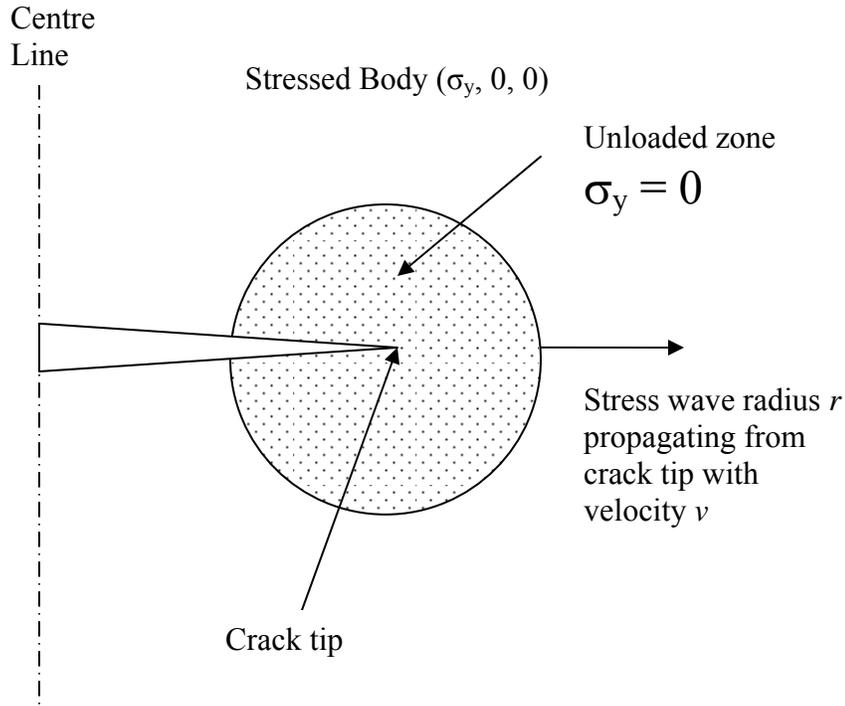

Fig. 4. Crack tip stress wave unloading model

Where the emerging cylindrical stress wave is frozen at a given instant in time, as defined below. This model is only applicable to a slow moving crack, for as shown by Baker (Baker 1962) for a fast moving crack the circular form of the stress wave propagation becomes elliptical with the waves in front of the crack becoming compressed, as photographed by Theocaris and Giorgiadis (Theocaris and Georgiadis 1984). Also as a brittle material is assumed the Rayleigh waves are ignored. This model is similar in concept to the J Contour Integral (Rice 1968). However, the J Contour Integral expresses the strain energy density at the crack tip by using a line integral, whereas this model estimates the energy released under the moving unloading wave front, and relates it to new crack surface generation.

The energy released from the body per unit thickness by the unloading crack tip stress wave, assuming that the crack opening is negligible, may be modelled as:

$$M' = \frac{\sigma_y^2}{2E}(vol) = \frac{\sigma_y^2}{2E}(\pi r^2) \tag{11}$$

Where $r$ is the radius of the unloading stress wave, and is a function of time, being dependent on the stress wave velocity $v$. However, at a given instant in time the local stress $\sigma_y$ is a function of $r$ and so taking the average stress as $\bar{\sigma}_y = \dfrac{K}{\sqrt{2\pi\bar{r}}}$ and taking $\bar{r} = a/2$ to correspond to the Griffith case, then the Lagrangian becomes:

$$L_i = 2\gamma a - \frac{K^2 a}{E} \tag{12}$$

So $\quad \dfrac{\partial L_i}{\partial a} = 2\gamma - \dfrac{K^2}{E} = 0 \tag{13}$

Hence $2\gamma E = K^2 = \sigma^2 \pi a \tag{14}$

Which is the Griffith-Irwin equation derived without recourse to two derivations defining the Griffith and Irwin criteria separately, and is a major simplification of the existing derivation.

It arises because by modelling the physics of the fracture process by a simplified model of the unloading of the body by a crack tip stress wave mechanism, it connects the energy released from the body with the K dominated stress field ahead of the crack. This approximation is much closer to the actual fracture mechanism, and gives a much clearer model of what is actually occurring in the fracture process of brittle materials.

In conclusion:

1) The monotonic Griffith's crack has been re-analysed using Hamiltonian mechanics, based on a modified Lagrangian, to express the energy transfer in the fracture process. The solutions obtained agree with Griffith's classical analysis.

2) A simplified crack tip stress wave unloading model has been introduced to describe the energy released from the body from the 'K' dominated stress field at the tip of the crack. The subsequent Lagrangian, and the solution from the Euler-Lagrange equation, are shown to give the Griffith's-Irwin solution in one step, without the need to invoke two separate derivations

3) It is believed that the depiction of the fracture process as given here is an approximate, but physically realistic, model of the fracture process, and considerably simplifies the derivation of the Griffith-Irwin relationship.